# Impact of carrier localization on recombination in InGaN quantum wells and the efficiency of nitride light-emitting diodes: insights from theory and numerical simulations

Christina M. Jones,[1] Chu-Hsiang Teng,[2] Qimin Yan,[3] Pei-Cheng Ku,[1,2] and Emmanouil Kioupakis[4,*]

[1] Applied Physics Program, University of Michigan, 450 Church St, Ann Arbor, MI, USA 48109
[2] Electrical Engineering and Computer Science Department, University of Michigan, 1301 Beal Ave, Ann Arbor, MI, USA 48109
[3] Department of Physics, Temple University, 1925 N. 12th St, Philadelphia, PA, USA 19122
[4] Materials Science and Engineering Department, University of Michigan, 2300 Hayward St, Ann Arbor, MI, USA 48109

We examine the effect of carrier localization due to random alloy fluctuations on the radiative and Auger recombination rates in InGaN quantum wells as a function of alloy composition, crystal orientation, carrier density, and temperature. Our results show that alloy fluctuations reduce individual transition matrix elements by the separate localization of electrons and holes, but this effect is overcompensated by the additional transitions enabled by translational symmetry breaking and the resulting lack of momentum conservation. Hence, we find that localization increases both radiative and Auger recombination rates, but that Auger recombination rates increase by one order of magnitude more than radiative rates. Furthermore, we demonstrate that localization has an overall detrimental effect on the efficiency-droop and green-gap problems of InGaN LEDs.

Indium gallium nitride (In$_x$Ga$_{1-x}$N) has transformed solid-state lighting by enabling efficient light-emitting diodes (LEDs)[1] and lasers[2] in the short-wavelength part of the visible spectrum. Despite their commercial success, InGaN devices suffer from several issues that may be further exacerbated by the localization of carriers. The internal quantum efficiency (IQE) of InGaN LEDs reduces at high power (efficiency droop),[3] which is further aggravated for LEDs operating at longer wavelengths (green-gap problem).[4] Auger recombination, a three-carrier scattering process that consumes the energy of a recombining electron-hole pair to excite another electron higher in the conduction band (electron-electron-hole Auger, or "eeh") or another hole lower in the valence band (hole-hole-electron Auger, or "hhe"), has been demonstrated to be an important non-radiative recombination mechanism in nitride LEDs and a plausible cause of the efficiency-droop and green-gap problems.[5-9]

Despite extensive studies on carrier localization in InGaN, its impact on radiative and Auger recombination remains unclear. Optoelectronic device simulations frequently use the virtual-crystal approximation (VCA), which models alloys as a crystalline solid with interpolated properties between the two end compounds. However, virtual crystals miss several physical features of disorder, such as translational symmetry breaking and the localization of carriers. InGaN alloys in particular contain statistically random composition fluctuations at the nanometer scale[10-12] that spatially localizes carriers[13-16]. Several studies have reported the effect of InGaN composition fluctuations on radiative and Auger recombination. Simulations by Yang *et al.* found radiative recombination rates in devices with fluctuating-alloy quantum wells (QWs) to be much *higher* than virtual-crystal alloys.[15] In contrast, Auf der Maur *et al.* used atomistic tight-binding and reported optical matrix elements and radiative recombination coefficients in fluctuating alloys as *smaller* than those of virtual crystals.[17] Experimentally, part of the reduction of the quantum efficiency of LEDs has also been attributed to hole localization.[18] With regards to Auger recombination, there are also conflicting results. Work using a semi-empirical model suggests that hole localization reduces Auger recombination,[19] but optical experiments indicate that localization increases Auger recombination.[20] Further studies are therefore necessary to understand the effects of alloy fluctuations on InGaN devices.

In this work, we use Schrödinger-Poisson simulations in the effective-mass approximation to investigate the effect of carrier localization on the radiative and Auger recombination rates in polar and non-polar InGaN QWs. We compare results for simulations incorporating alloy composition fluctuations to virtual-crystal InGaN QWs and to virtual-crystal bulk InGaN to assess the modification of recombination rates by localization. We find that carrier localization increases Auger recombination rates by one order of magnitude more than radiative rates, and is therefore detrimental to the IQE of InGaN LEDs and lasers.

We first demonstrate that the radiative and Auger recombination rates of localized carriers in InGaN are approximately proportional to the weighted-averaged overlaps squared of the localized wave functions of the recombining carriers. The detailed derivation (provided in the supplementary material) is similar to the analysis for one-dimensional confinement in QWs.[9] The ratio of radiative coefficients of fluctuating or virtual-crystal alloys compared to bulk virtual crystals is determined by

$$\frac{B^{\text{fluct/VCA,QW}}}{B^{\text{VCA,bulk}}} = \frac{\overline{F_{eh}^2}^{\text{fluct/VCA,QW}}}{\overline{F_{eh}^2}^{\text{VCA,bulk}}}, \quad (1)$$

where

$$\overline{F_{eh}^2} \equiv \frac{\sum_{1,2} f_1 (1-f_2) |\int \varphi_1^*(r)\varphi_2(r) dr|^2}{\sum_{1,2} f_1 (1-f_2)} \quad (2)$$

are weighted-averaged overlaps squared of localized electron and hole envelope functions $\varphi$, weighted by the Fermi-Dirac occupation numbers $f$. We subsequently demonstrate that the ratio of Auger coefficients of fluctuating (virtual crystal) alloys versus bulk virtual crystals is given by

$$\frac{C_{eeh}^{\text{fluct/VCA,QW}}}{C_{eeh}^{\text{VCA,bulk}}} = \frac{\overline{F_{eeh}^2}^{\text{fluct/VCA,QW}}}{\overline{F_{eeh}^2}^{\text{VCA,bulk}}} \quad (3)$$

for the case of eeh Auger recombination, where

$$\overline{F_{eeh}^2} \equiv \frac{\sum_{1,2,3} f_1 f_2 (1-f_3) V_{QW} |\int \varphi_1^*(r)\varphi_2^*(r)\varphi_3(r) dr|^2}{\sum_{1,2,3} f_1 f_2 (1-f_3)} \quad (4)$$

are weighted-averaged triple overlaps squared of localized electron and hole envelope functions (and similarly for hhe Auger). Polkovnikov and Zegrya derived a similar expression for Auger recombination in QWs.[21] Shockley-Read-Hall rates are also approximately proportional to $\overline{F_{eh}^2}$ by similar arguments as discussed in Ref. 9 for localization along the QW confinement direction. Parameters for the carrier-density and temperature-dependent bulk $B$ and $C$ coefficients are obtained from density functional theory (DFT) results.[22,23] For a carrier density of $10^{19}$ cm$^{-3}$ and a temperature of 300 K (typical LED operating conditions) the bulk coefficients are $B_0 = 2\times10^{-11}$ cm$^3$/s, $C_{eeh,0} = 2\times10^{-31}$ cm$^6$/s, and $C_{hhe,0} = 4\times10^{-31}$ cm$^6$/s.[5,23] DFT calculations include alloy disorder but the simulation cells are too small (32 atoms) to exhibit localization of carriers, and thus are approximately closer to VCA results. For the bulk Shockley-Read-Hall coefficient we used a typical experimental value ($6\times10^{7}$ s$^{-1}$).[5]

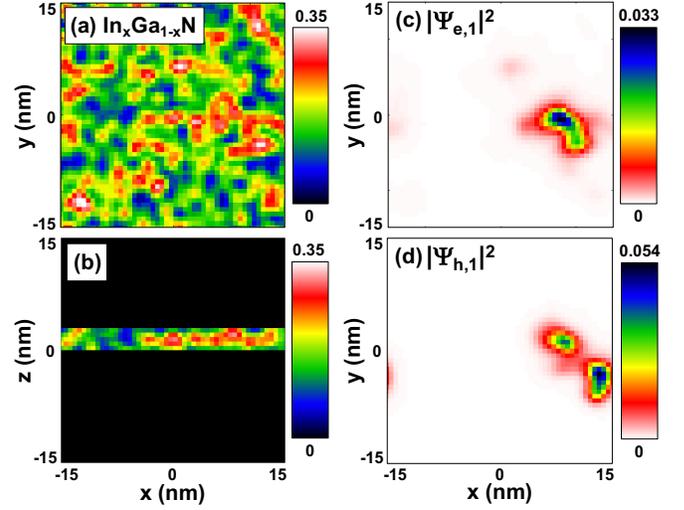

FIG. 1. (a,b) A randomly generated, local InGaN distribution in a 3 nm nominal $In_{0.15}Ga_{0.85}N$ polar quantum well ($z \parallel c$) at (a) the x-y plane (c-plane) at the QW center (z = 1.5 nm) and (b) a cross-section in the x-z plane (y = 0). Square modulus of electron (c; z = 2.3 nm) and hole (d; z = 0) wave functions for the alloy distribution in (a,b).

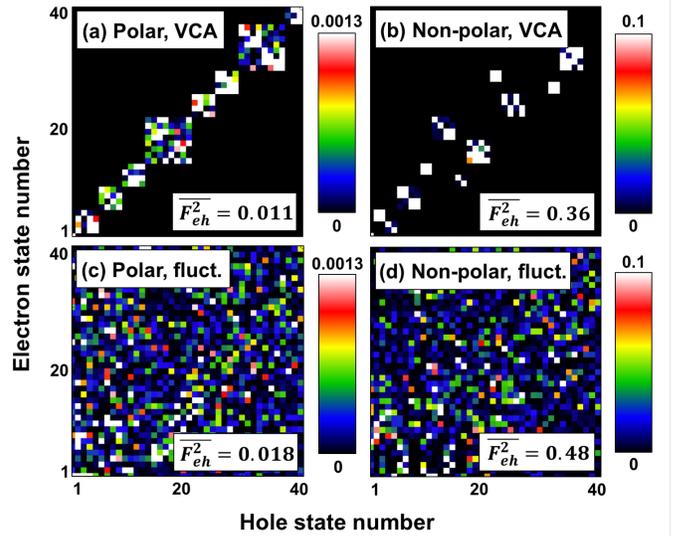

FIG. 2. Individual electron-hole wave-function overlaps between the first 40 electron and 40 hole states, illustrating the symmetry breaking introduced by alloy fluctuations. (a,b) Wave-function overlaps for a virtual-crystal alloy in the (a) polar and (b) non-polar orientation. (c,d) Wave-function overlaps for a fluctuating alloy in the (c) polar and (d) non-polar orientation. The weighted-averaged matrix element is larger for the fluctuating alloy because the reduction of individual matrix elements by spatial separation is overcompensated by the larger number of allowed electron-hole combinations due to translational symmetry breaking.

We subsequently numerically evaluate wave functions and energies for realistic 3 nm InGaN QWs with a commercial Schrödinger-Poisson solver (nextnano[24]). We consider the effect of the occupied states (for 99% of the carriers) in the effective-mass approximation for all heavy hole, light hole, and crystal-field split-off bands, which are all occupied by carriers and have been found to be important in, e.g., Ref. 25. For In$_x$Ga$_{1-x}$N compositions spanning the blue-green spectrum (10-35%), we simulated ten random 3D alloy distributions using the method described in Ref. 15 (see supplementary material for more information). Fermi levels and state occupancies are evaluated in the rigid-band approximation (discussed further in supplementary material). Our main results are evaluated with a spatial averaging of the alloy composition and a simulation grid spacing of 0.6 nm, which is smaller than the localization length of 1-3 nm (5-10 nm) for holes (electrons) reported for InGaN.[16,26] We validated the conclusions of our study with calculations on a finer grid (details below and in supplementary material). We examine two QW polarization extremes: the polar, $c$-plane and the non-polar, $a$-plane (11$\underline{2}$0) orientation. These two polarizations also approximate no screening and complete screening of polarization by carriers, respectively. Realistic devices with externally applied voltages and screening by free carriers lie between these two polar and non-polar extremes. Our study therefore enables us to separate the effect of polarization and localization on carrier overlaps and recombination rates.

We then evaluate the weighted-averaged double and triple wave-function overlaps to quantify the impact of localization on the recombination rates. Our results show that both electrons and holes are localized by composition fluctuations (Fig. 1), similar to published work.[15-17] Overlaps for virtual crystals obey selection rules, and only electron and holes states of the same wave vector demonstrate nonzero overlap. However, random alloy fluctuations and carrier localization break the translational symmetry and wave-vector conservation, allowing nonzero overlaps between any hole and electron states. These nonzero overlaps are typically much smaller than 1 (Fig. 2) since carriers localize in different spatial regions. Overall, however, by including the contribution from all occupied states, localization increases the electron and hole weighted overlaps compared to virtual crystals (Fig. 2). For In$_{0.15}$Ga$_{0.85}$N QWs, and for a carrier density of $10^{19}$ cm$^{-3}$ and a temperature of 300 K, alloy fluctuations increase the electron-hole weighted-averaged overlap by 64% (31%) over the virtual crystal in the polar (non-polar) orientation.

Our calculations demonstrate that both the radiative and the Auger coefficients are larger in fluctuating alloys than in virtual crystals; however, Auger coefficients increase significantly more than radiative coefficients (Fig. 3), indicating an overall detrimental effect of carrier localization on the LED IQE. All ten generated random alloy distributions were used for calculating $B$, but only two of these distributions were used for calculating $C$ because the triple wave-function overlaps are computationally intensive. The two distributions were those that gave $B$ values closest to the ten-distribution average and thus best represent the randomness of the alloy, which likely explains why the $C$ values for a given indium concentration have a smaller relative spread than the $B$ values. For a carrier density of $10^{19}$ cm$^{-3}$ and a temperature of 300 K, we observe a 30-50% increase in $B$ and 390-540% increase in $C$ in fluctuating alloys compared to virtual crystals.

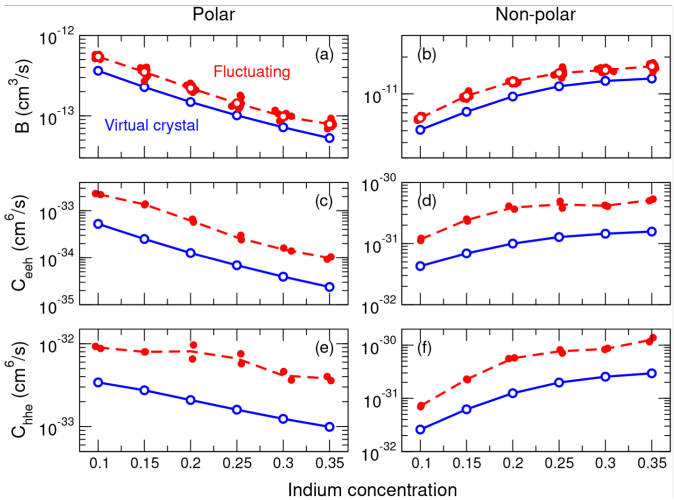

FIG. 3. Radiative ($B$) and Auger ($C_{eeh}$, $C_{hhe}$) coefficients of InGaN QWs calculated as a function of indium concentration for polar and non-polar orientations ($10^{19}$ cm$^{-3}$ carrier density, 300 K carrier temperature). In all cases, alloy composition fluctuations increase the coefficients compared to virtual-crystal alloys.

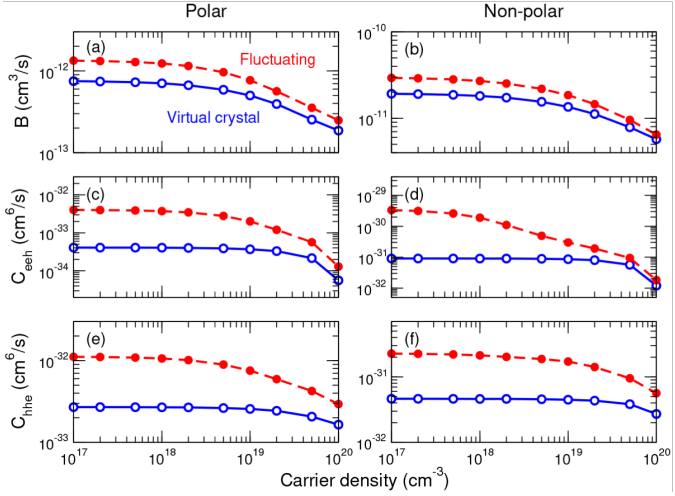
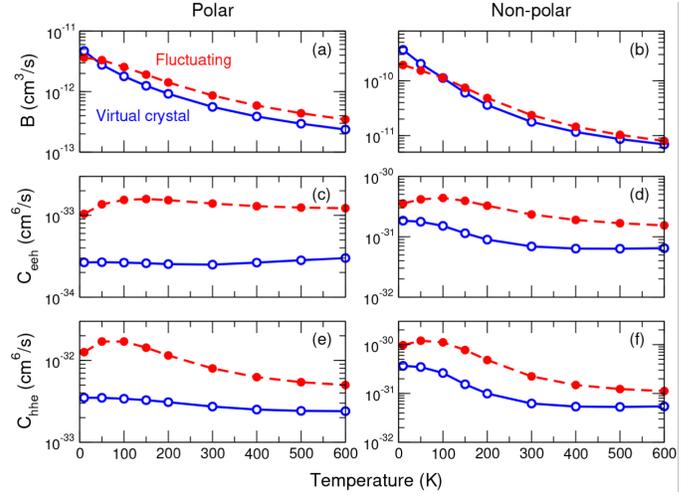

FIG. 4. Radiative ($B$) and Auger coefficients ($C_{eeh}$, $C_{hhe}$) of an In$_{0.15}$Ga$_{0.85}$N QW as a function of carrier density (300 K carrier temperature). In all cases, composition fluctuations increase Auger coefficients more than the corresponding radiative coefficients.

FIG. 5. Radiative ($B$) and Auger coefficients ($C_{eeh}$, $C_{hhe}$) of an In$_{0.15}$Ga$_{0.85}$N QW as a function of carrier temperature ($10^{19}$ cm$^{-3}$ carrier density). Except for very low temperatures, Auger coefficients of fluctuating alloys increase over those of virtual crystals by larger factors than radiative coefficients.

To facilitate comparison with atomistic tight-binding calculations, we simulated a In$_{0.15}$Ga$_{0.85}$N fluctuating alloy with a 0.3 nm grid size, which is similar to the in-plane lattice constant of InGaN, and performed no spatial composition averaging. We found qualitatively identical results to our calculations with the 0.6 nm grid and spatial averaging: Auger coefficients increase significantly more than radiative coefficients due to fluctuations. In fact, we find that the fine-grid results demonstrate an even larger reduction in efficiency due to alloy fluctuations than the coarser-grid study (details in the supplementary material).

We also evaluate the radiative and Auger coefficients over a realistic range ($10^{17}$-$10^{20}$ cm$^{-3}$) of carrier densities for LED operation for the alloy distribution producing the $B$ value closest to the average $B$ value for the ten initial distributions. Compared to virtual crystals, we find that composition fluctuations consistently increase Auger coefficients more than the radiative ones for all examined carrier densities (Fig. 4). At low carrier concentrations, the enhancement of fluctuating-alloy coefficients over virtual crystals is particularly prominent because localization effects are strongest for the lowest-energy states that dominate state occupancies. At high carrier densities, state occupancies are dominated by less confined states and fluctuating-alloy coefficients approach those of virtual crystals. The lower smoothness of the Auger curves (Figs. 3(c)-(f)) compared to the radiative ones (Figs. 3(a) and (b)) is likely due to statistical limitations and could be improved with additional calculations for more random alloy distributions and larger simulation cells.

We note that $C_{hhe}$ is larger than $C_{eeh}$ for polar QWs, in agreement with bulk DFT calculations.[23] Interestingly, for non-polar QWs (and, approximately, screened polar QWs), we find $C_{eeh}$ to be larger than $C_{hhe}$ by a factor of 1–14, depending on the carrier density, which agrees with a recent experimental report that $C_{eeh}$ is larger than $C_{hhe}$ by a factor of 1–12.[27] Our results therefore reveal that the interplay of QW composition, carrier localization, polarization fields, quantum confinement, carrier density, and screening plays an important role in determining the dominant Auger recombination type (eeh or hhe) in InGaN QWs.

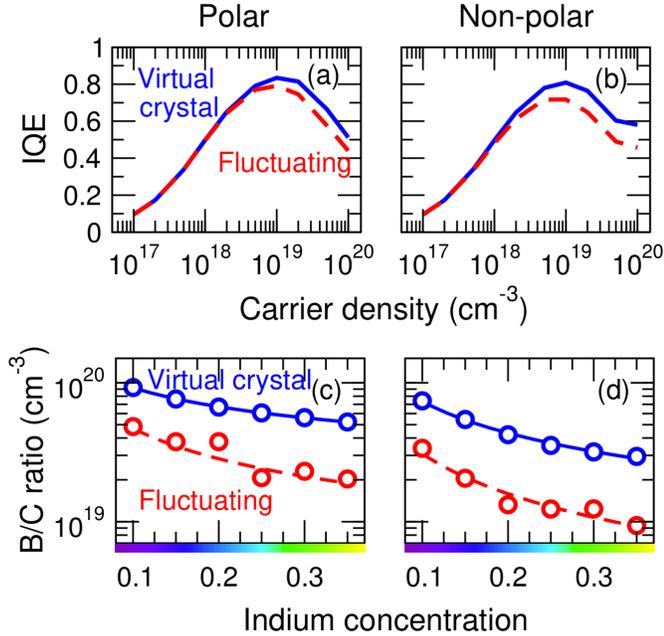

FIG. 6. (a,b) Internal quantum efficiency (IQE) of $In_{0.15}Ga_{0.85}N$ quantum wells with fluctuating alloys (dashed line) and virtual crystals (solid line). Fluctuating alloys exhibit lower IQE than virtual crystals. (c,d) Ratio of radiative ($B$) to Auger ($C$) coefficients as a function of indium concentration (300 K carrier temperature, $10^{19}$ cm$^{-3}$ carrier density). The ratio decreases with indium concentration faster for fluctuating alloys than for virtual crystals, consequently alloy fluctuations exacerbate the green-gap problem.

We further determine the temperature dependence of the radiative and Auger coefficients in fluctuating alloys and compare to virtual crystals. As with the carrier density study, for the temperature study we use the alloy distribution that produced the B value closest to the average of the ten initial distributions. Across a temperature range of 100–600 K, Auger coefficients increase more than the radiative ones due to alloy fluctuations (Fig. 5). Auger coefficients for fluctuating alloys increase with increasing temperature up to 100 K, at which point they start to decrease. This peak with temperature was observed experimentally in GaInP alloys, which also exhibit carrier localization.[28] At low temperatures (below 100 K), carriers are strongly localized and wave-function overlaps are small, therefore the spatial separation of carriers dominates and reduces the radiative coefficients compared to virtual crystals. Above 100 K, the occupations of higher-energy extended states increases with increasing temperature, and localized states contribute a smaller fraction to the weighted-averaged overlaps, hence the coefficients decrease with temperature and approach VCA results.

Our analysis resolves the apparent disagreement between published studies on the impact of carrier localization on the radiative and Auger rates. Our results agree with other reported simulations in that for the ground state and other individual state overlaps, fluctuations reduce the overlaps compared to virtual crystals.[17,19] However, by considering the occupations of all electron and hole states for realistic carrier densities and temperatures under device-operating conditions, we predict an overall net increase of the radiative recombination coefficients. Our results are therefore consistent with the work of Yang *et al.* and Shahmohammadi *et al.*, who found localization to increase recombination rates.[15,20]

Our results further demonstrate that carrier localization exacerbates the efficiency-droop and green-gap problems of both polar and non-polar LEDs. The difference between the IQE of fluctuating alloys and virtual crystals is up to 9% (12%) in the polar (non-polar) orientation (Fig. 6). Additionally, the $B/C$ ratio decreases more with increasing indium concentration for fluctuating alloys than for virtual crystals, indicating that alloy fluctuations aggravate the green-gap problem (Fig. 6). Random composition fluctuations are an inevitable result of alloying, therefore the detrimental effects of carrier localization on the efficiency droop of InGaN LEDs cannot be eliminated. One avenue to avoid localization is to use binary nitride compounds and tune the gap by engineering the polarization fields or the quantum-confinement dimensions.[29,30] Another promising avenue to mitigate Auger recombination is to reduce the steady-state carrier density with fewer, thicker InGaN QWs (e.g., a double-heterostructure design up to 12 nm thick).[6,31]

In conclusion, we quantify the effect of carrier localization due to alloy fluctuations on the IQE of InGaN QWs. Localization increases the Auger rates more than radiative rates, exacerbating the efficiency-droop and green-gap problems of InGaN LEDs.

See supplementary material (below) for more details on our computational methods, the formalism of recombination for localized wave functions, details on validating our alloy distribution methods with fine-grid studies, and discussions of excitons and alloy composition characterization.

We thank Stefan Birner, Max Falkowski, Alvaro Gomez-Iglesias, and Sergey Karpov for useful discussions. This work was supported by the National Science Foundation through Grant Nos. DMR-1254314 (Schrödinger-Poisson and recombination-rate calculations) and DMR-1409529 (alloy microstructure simulations). C. M. Jones acknowledges support from the National Science Foundation Graduate Research Fellowship Program through Grant No. DGE-1256260. This research used resources of the National Energy Research Scientific Computing (NERSC) Center, a DOE Office of Science User Facility supported under Contract No. DE-AC02-05CH11231.

This work was published in Applied Physics Letters:

C.M. Jones, C.-H. Teng, Q. Yan, P.-C. Ku, and E. Kioupakis. Appl. Phys. Lett. **111**, 113501 (2017); doi: 10.1063/1.5002104.

Available online: http://dx.doi.org/10.1063/1.5002104.

[1] S. Nakamura, T. Mukai, and M. Senoh, Appl. Phys. Lett. **64**, 1687 (1994).
[2] S. Nakamura, M. Senoh, S.-I. Nagahama, N. Iwasa, T. Yamada, T. Matsushita, H. Kiyoku, and Y. Sugimoto, Jpn. J. Appl. Phys. **35**, L217 (1996).
[3] J. Piprek, Phys. Stat. Sol. (a) **207**, 2217 (2010).
[4] C.J. Humphreys, MRS Bull. **33**, 459 (2011).
[5] Y.C. Shen, G.O. Mueller, S. Watanabe, N.F. Gardner, A. Munkholm, and M.R. Krames, Appl. Phys. Lett. **91**, 141101 (2007).
[6] N.F. Gardner, G.O. Müller, Y.C. Shen, G. Chen, S. Watanabe, W. Götz, and M.R. Krames, Appl. Phys. Lett. **91**, 243506 (2007).
[7] A. Laubsch, M. Sabathil, J. Baur, M. Peter, and B. Hahn, IEEE Trans. Electron Devices **57**, 79 (2009).
[8] E. Kioupakis, P. Rinke, K.T. Delaney, and C.G. Van de Walle, Appl. Phys. Lett. **98**, 161107 (2011).
[9] E. Kioupakis, Q. Yan, and C.G. Van de Walle, Appl. Phys. Lett. **101**, 231107 (2012).
[10] M.J. Galtrey, R.A. Oliver, M.J. Kappers, C.J. Humphreys, P.H. Clifton, D. Larson, D.W. Saxey, and A. Cerezo, J. Appl. Phys. **104**, 013524 (2008).
[11] S.E. Bennett, D.W. Saxey, M.J. Kappers, J.S. Barnard, C.J. Humphreys, G.D. Smith, and R.A. Oliver, Appl. Phys. Lett. **99**, 021906 (2011).
[12] Y.-R. Wu, R. Shivaraman, K.-C. Wang, and J.S. Speck, Appl. Phys. Lett. **101**, 083505 (2012).
[13] P.R.C. Kent and A. Zunger, Appl. Phys. Lett. **79**, 1977 (2001).
[14] Z. Li, J. Kang, B. Wei Wang, H. Li, Y. Hsiang Weng, Y.-C. Lee, Z. Liu, X. Yi, Z. Chuan Feng, and G. Wang, J. Appl. Phys. **115**, 083112 (2014).
[15] T.-J. Yang, R. Shivaraman, J.S. Speck, and Y.-R. Wu, J. Appl. Phys. **116**, 113104 (2014).
[16] S. Schulz, M.A. Caro, C. Coughlan, and E.P. O'Reilly, Phys. Rev. B **91**, 035439 (2015).
[17] M. Auf der Maur, A. Pecchia, G. Penazzi, W. Rodrigues, and A. Di Carlo, Phys. Rev. Lett. **116**, 027401 (2016).
[18] F. Nippert, S.Y. Karpov, G. Callsen, B. Galler, T. Kure, C. Nenstiel, M.R. Wagner, M. Strassburg, H.-J. Lugauer, and A. Hoffmann, Appl. Phys. Lett. **109**, 161103 (2016).
[19] S.Y. Karpov, Photon. Res. **5**, A7 (2017).
[20] M. Shahmohammadi, W. Liu, G. Rossbach, L. Lahourcade, A. Dussaigne, C. Bougerol, R. Butté, N. Grandjean, B. Deveaud, and G. Jacopin, Phys. Rev. B **95**, 125314 (2017).
[21] A.S. Polkovnikov and G.G. Zegrya, Phys. Rev. B **58**, 4039 (1998).
[22] E. Kioupakis, Q. Yan, D. Steiauf, and C.G. Van de Walle, New J. Phys. **15**, 125006 (2013).
[23] E. Kioupakis, D. Steiauf, P. Rinke, K.T. Delaney, and C.G. Van de Walle, Phys. Rev. B **92**, 035207 (2015).
[24] S. Birner, T. Zibold, T. Andlauer, T. Kubis, M. Sabathil, A. Trellakis, and P. Vogl, IEEE Trans. Electron Devices **54**, 2137 (2007).
[25] G. Rossbach, J. Levrat, G. Jacopin, M. Shahmohammadi, J.-F. Carlin, J.D. Ganière, R. Butté, B. Deveaud, and N. Grandjean, Phys. Rev. B **90**, 201308 (2014).
[26] D. Watson-Parris, M.J. Godfrey, P. Dawson, R.A. Oliver, M.J. Galtrey, M.J. Kappers, and C.J. Humphreys,

Phys. Rev. B **83**, 115321 (2011).

[27] A. Nirschl, M. Binder, M. Schmid, I. Pietzonka, H.-J. Lugauer, R. Zeisel, M. Sabathil, D. Bougeard, and B. Galler, Opt. Express **24**, 2971 (2016).

[28] Z.C. Su, J.Q. Ning, Z. Deng, X.H. Wang, S.J. Xu, R.X. Wang, S.L. Lu, J.R. Dong, and H. Yang, Nanoscale **8**, 7113 (2016).

[29] D. Bayerl and E. Kioupakis, Nano Lett. **14**, 3709 (2014).

[30] D. Bayerl, S.M. Islam, C.M. Jones, V. Protasenko, D. Jena, and E. Kioupakis, Appl. Phys. Lett. **109**, 241102 (2016).

[31] S. Okur, M. Nami, A.K. Rishinaramangalam, S.H. Oh, S.P. DenBaars, S. Liu, I. Brener, and D.F. Feezell, Opt. Express **25**, 2178 (2017).

# Supplementary Material

## I. ADDITIONAL DETAILS ON COMPUTATIONAL METHODS

Since several studies have reported the distribution of indium and gallium atoms in InGaN to be statistically random,[1-4] we use a random-number generator to assign the atomic occupancies (In or Ga) on a real-space point grid according to a binary (Bernoulli) distribution based on the nominal indium composition. We subsequently average the composition in the volume ($\pm$ 1.2 nm along the x, y, and z directions) surrounding each grid point to better approximate atom probe tomography (APT) results. The resulting alloy distributions reproduce the statistical distributions of APT data and follow binomial composition distributions.[2,5] Indium clusters form solely due to random composition fluctuations. We also incorporated a Gaussian out-of-plane distribution profile.[6]

Fermi levels and state occupancies are evaluated in the rigid-band approximation. The rigid-band approximation ignores carrier interactions and assumes that carrier energies and wave functions are not affected by carrier occupations, i.e., the conduction/valence structure of doped or excited semiconductors is the same as the un-doped ground-state one. The validity of the approximation has been demonstrated in, e.g., carrier transport in thermoelectric materials.[7] Interactions between localized holes may affect their electronic structure, but we expect their localization properties to remain similar upon the inclusion of interactions.

The effective mass approximation has been validated for nanostructures and local fluctuations in previous work.[1,8,9]

## II. THEORY OF RECOMBINATION FOR LOCALIZED WAVE FUNCTIONS

Here we derive the equations that relate the radiative and Auger coefficients of fluctuating alloys to the corresponding coefficients of virtual-crystal alloys and weighted-averaged overlaps squared of envelope functions. We start by expressing the localized wave functions of fluctuating alloys as a linear combination of virtual-crystal bulk Bloch functions:

$$\psi_J(\boldsymbol{r}) = \frac{1}{\sqrt{V}}\sum_{\boldsymbol{k}} c_J(\boldsymbol{k}) e^{i\boldsymbol{k}\cdot\boldsymbol{r}} u_{n\boldsymbol{k}}(\boldsymbol{r}) \cong \varphi_J(\boldsymbol{r}) u_n(\boldsymbol{r}), \tag{S1}$$

where the periodic part of the Bloch wave function can approximately be considered independent of the wave vector,

$$u_{n\boldsymbol{k}}(\boldsymbol{r}) \cong u_n(\boldsymbol{r}), \tag{S2}$$

and the envelope function of the localized carriers is given by,

$$\varphi_J(\boldsymbol{r}) = \frac{1}{\sqrt{V}}\sum_{\boldsymbol{k}} c_J(\boldsymbol{k}) e^{i\boldsymbol{k}\cdot\boldsymbol{r}}, \tag{S3}$$

where the composite index,

$$\boldsymbol{J} \equiv (j, n), \tag{S4}$$

is a combination of the index of the localized state $j$ and the bulk band index $n$. In this basis, the radiative recombination rate (i.e., the number of recombining carriers per unit time per unit volume) is determined by Fermi's golden rule according to:

$$R_{\text{rad}} = 2\frac{2\pi}{\hbar}\sum_{1,2} f_1(1-f_2)|\langle 1|H_{\text{el-phot}}|2\rangle|^2 \delta(\epsilon_1 - \epsilon_2 - \hbar\omega), \tag{S5}$$

where the matrix elements of the electron-photon perturbation Hamiltonian between localized states can be expressed in the basis of delocalized Bloch orbitals as:

$$\langle 1|H_{\text{el-phot}}|2\rangle = \sum_{k_1,k_2} c_1^*(k_1) c_2(k_2) \langle n_1 k_1|H_{\text{el-phot}}|n_2 k_2\rangle. \tag{S6}$$

In bulk virtual-crystal alloys, the translational crystalline symmetry conserves the crystal momentum in radiative transitions, and the matrix elements are given by:

$$\langle n_1 k_1|H_{\text{el-phot}}|n_2 k_2\rangle \propto p_{n_1 n_2}(k_1)\delta_{k_1,k_2} \cong p_{n_1 n_2}\delta_{k_1,k_2}, \tag{S7}$$

where the interband matrix elements are further assumed independent of wave vector. In this case, the electron-photon matrix elements between localized electrons and holes are proportional to the electron and hole envelope function overlap according to:

$$\langle 1|H_{\text{el-phot}}|2\rangle \propto p_{n_1 n_2} \sum_{k_1} c_1^*(k_1) c_2(k_1) \propto p_{n_1 n_2} \int \varphi_1^*(r)\varphi_2(r) dr. \tag{S8}$$

Therefore, the radiative recombination rate is given by:

$$R_{\text{rad}} = Bnp \propto \sum_{1,2} f_1(1-f_2)|\int \varphi_1^*(r)\varphi_2(r) dr|^2, \tag{S9}$$

i.e., it is proportional to the weighted sum of electron-hole envelope function overlap squared, the weights being the electron and hole Fermi-Dirac occupation factors. The radiative coefficient is subsequently given by

$$B \propto \frac{\sum_{1,2} f_1(1-f_2)|\int \varphi_1^*(r)\varphi_2(r) dr|^2}{\sum_{1,2} f_1(1-f_2)} \equiv \overline{F_{eh}^2}, \tag{S10}$$

i.e., it is proportional to the weighted-averaged overlap squared of electron and hole wave functions. The last equation allows the comparison of radiative recombination coefficients of virtual-crystal and fluctuating-alloy quantum wells to bulk virtual crystals according to:

$$\boxed{\frac{B^{\text{fluct/VCA,QW}}}{B^{\text{VCA,Bulk}}} = \frac{\overline{F_{eh}^2}^{\text{fluct/VCA,QW}}}{\overline{F_{eh}^2}^{\text{VCA,Bulk}}}}. \tag{S11}$$

We extend the previous analysis to encompass Auger recombination. We examine electron-electron-hole recombination (the hole-hole-electron case is analogous). The recombination rate is given by Fermi's golden rule according to:

$$R_{\text{Auger,eeh}} = 2\frac{2\pi}{\hbar}\sum_{1,2,3,4} f_1 f_2 (1-f_3)(1-f_4)|\langle 12|H_{\text{Auger}}|34\rangle|^2 \delta(\epsilon_1 + \epsilon_2 - \epsilon_3 - \epsilon_4), \tag{S12}$$

where the Auger recombination matrix elements are the screened Coulomb interaction matrix elements between the initial and final pairs of states. Expressed in the Bloch basis of virtual crystals, the matrix elements are given by:

$$\langle 12|H_{\text{Auger}}|34\rangle = \iint \psi_1^*(r)\psi_2^*(r')W(r,r')\psi_3(r)\psi_4(r') dr dr'$$

$$= \sum_{k_1,k_2,k_3,k_4} c_1^*(k_1) c_2^*(k_2) c_3(k_3) c_4(k_4) \langle n_1 k_1 n_2 k_2 | W | n_3 k_3 n_4 k_4 \rangle. \tag{S13}$$

In general, the Coulomb matrix elements are nontrivial functions of the wave vectors. They involve the wave-vector dependence of the screened Coulomb interaction, as well as the wave-vector dependence of the overlaps between the periodic parts of Bloch functions across bands.[10] However, the matrix elements can be simplified in the wide-band-gap group-III nitrides. Since the energy released by the recombining electron-hole pair is large, on the order of 2.4-3.0 eV for InGaN used for green to violet LEDs, the resulting Auger electron is excited to high conduction-band states at energies higher than the conduction-band minimum by approximately the value of the gap. The momentum transfer involved for these high-energy Auger transitions is large, which indicates that Coulomb scattering in wide-gap-nitrides is short-ranged.[11] As demonstrated in Fig. S1, the dominant Auger matrix elements in group-III nitrides are approximately constant and independent of wave vector at short range (large momentum transfer).

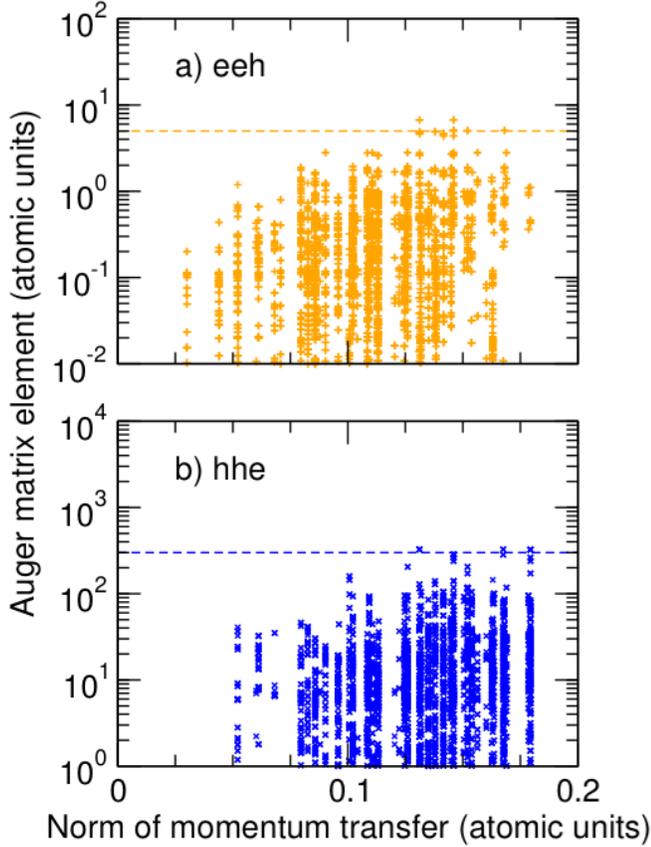

FIG. S1. Auger matrix elements of GaN for (a) electron-electron-hole (eeh) and (b) hole-hole-electron (hhe) Auger recombination as a function of the momentum transfer involved in the Auger process. For short-range Auger scattering (i.e., large momentum transfer), which dominates Auger recombination in wide-band-gap nitrides, the dominant matrix elements reach values that are approximately independent of momentum transfer (indicated with dotted line).

We can therefore approximately express the Coulomb matrix elements between virtual-crystal states as a constant times a Kronecker delta for momentum conservation:

$$\langle n_1 k_1 n_2 k_2 | W | n_3 k_3 n_4 k_4 \rangle \cong W \delta_{k_1+k_2-k_3-k_4}. \tag{S14}$$

Under the above approximation, valid for other wide-band-gap materials as well, we can write the Auger matrix elements in terms of the envelope functions as:

$$\langle 12 | H_{\text{Auger}} | 34 \rangle \cong W \sum_{k_1,k_2,k_3} c_1^*(k_1) c_2^*(k_2) c_3(k_3) c_4(k_1+k_2-k_3)$$
$$= W \sum_{k_1,k_2,k_3} \int \varphi_1^*(r_1) e^{ik_1 \cdot r_1} \varphi_2^*(r_2) e^{ik_2 \cdot r_2} \varphi_3(r_3) e^{-ik_3 \cdot r_3} \varphi_4(r_4) e^{-i(k_1+k_2-k_3) \cdot r_4} dr_1 dr_2 dr_3 dr_4. \tag{S15}$$

Now, we make use of the identity:

$$\sum_{k_1} e^{ik_1 \cdot (r_1 - r_4)} = V\delta(r_1 - r_4) \tag{S16}$$

to reduce the integral to a simple overlap integral of the four envelope wave functions of the states involved in the Auger transition:

$$\langle 12|H_{\text{Auger}}|34\rangle \propto \int \varphi_1^*(r)\varphi_2^*(r)\varphi_3(r)\varphi_4(r)dr. \tag{S17}$$

Now, state **4** is highly excited and its energy significantly exceeds the localization energy. It is therefore highly delocalized over the entire volume of the material, and its envelope function can be approximated to be constant. The Auger matrix elements can therefore be reduced to triple overlap integrals according to:

$$\langle 12|H_{\text{Auger}}|34\rangle \propto \int \varphi_1^*(r)\varphi_2^*(r)\varphi_3(r)dr. \tag{S18}$$

Correspondingly, the Auger recombination rate is given by a weighted sum of the squares of triple overlaps, the weights given by the occupation numbers of the recombining electrons and holes according to:

$$R_{\text{Auger,eeh}} = C_{\text{eeh}} n^2 p \propto \sum_{1,2,3} f_1 f_2 (1-f_3) V_{QW} |\int \varphi_1^*(r)\varphi_2^*(r)\varphi_3(r)dr|^2, \tag{S19}$$

where $V_{QW}$ is the volume of the quantum well, and thus the Auger coefficients are proportional to weighted-averaged triple overlaps squared according to:

$$C_{\text{eeh}} \propto \frac{\sum_{1,2,3} f_1 f_2 (1-f_3) V_{QW} |\int \varphi_1^*(r)\varphi_2^*(r)\varphi_3(r)dr|^2}{\sum_{1,2,3} f_1 f_2 (1-f_3)} \equiv \overline{F_{\text{eeh}}^2}. \tag{S20}$$

By evaluating the weighted-averaged triple overlaps squared of fluctuating alloys and comparing to virtual crystals we can determine the impact of carrier localization on the Auger coefficients according to:

$$\boxed{\frac{C_{\text{eeh}}^{\text{fluct/VCA,QW}}}{C_{\text{eeh}}^{\text{VCA,bulk}}} = \frac{\overline{F_{\text{eeh}}^2}^{\text{fluct/VCA,QW}}}{\overline{F_{\text{eeh}}^2}^{\text{VCA,bulk}}}} \tag{S21}$$

and similarly for the hole-hole-electron process:

$$\boxed{\frac{C_{hhe}^{\text{fluct/VCA,QW}}}{C_{hhe}^{\text{VCA,bulk}}} = \frac{\overline{F_{hhe}^2}^{\text{fluct/VCA,QW}}}{\overline{F_{hhe}^2}^{\text{VCA,bulk}}}}. \tag{S22}$$

### III. STUDY USING FINE GRID AND ROUGH ALLOY COMPOSITION

We performed a set of calculations for a fluctuating alloy on a finer simulation mesh that closer compares to atomistic, tight-binding calculations. We use a fluctuating In$_{0.15}$Ga$_{0.85}$N alloy with a smaller grid size than the main part of our study (0.3 nm instead of 0.6 nm), which is similar to the in-plane lattice constants of InGaN. Additionally, the fine-grid calculations do not spatially average the alloy distribution in order to represent potential fluctuations at the atomic scale (Fig. S2). Results for the finer grid with no spatial averaging validate the physics and qualitative trends observed in the main part of our study, and maintain that composition fluctuations increase the Auger coefficients significantly more than the radiative ones. Additionally, the fine-grid results show an even larger quantitative internal quantum efficiency reduction due to alloy fluctuations than those observed with the coarse grid.

All wave functions obtained with the fine-grid calculations exhibit similar localization behavior to the coarse-grid study, except for some of the lowest-energy states. The first few electron states of the coarse-grid study, which cover the lowest 15-20 meV of electron energy states, demonstrate localization by compositions fluctuations. Above this energy, electron states are still affected by alloy fluctuations, but they become extended and cover the span of the simulation domain, while their amplitude decreases as expected for more extended states. In comparison, electron wave functions for the fine-grid study are also influenced by the alloy distribution, but they are extended even for the lowest-energy states and do not exhibit confinement. However, only the first three electron states out of the >45 states exhibit this different electron localization behavior between the coarse- and fine-grid studies. Therefore, the large majority of electrons are not affected by localization at low energies for the coarse-grid, spatially averaged alloy study, similar to the fine-grid results.

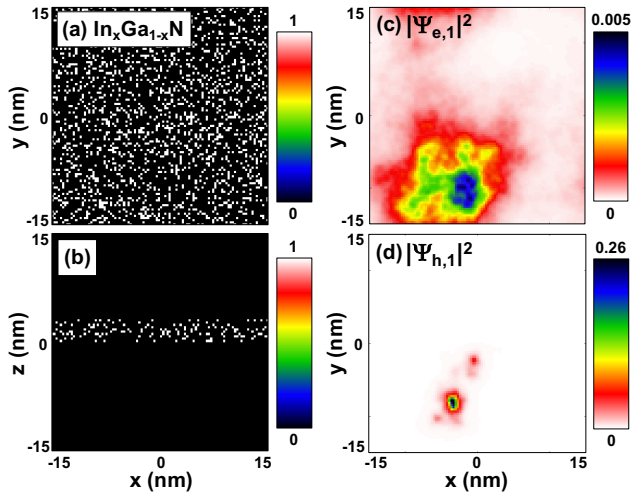

FIG. S2. (a,b) A random alloy distribution for a 3 nm polar $In_{0.15}Ga_{0.85}N$ QW, representing a more atomistic approach (0.3 nm grid and no spatial averaging). (a) shows the x-y plane at the QW center, $z = 1.5$ nm, and (b) shows a cross-section at $y = 0$. (c,d) The square modulus of electron and hole ground-state wave functions, respectively, for the random alloy distribution shown in (a,b). Each wave function is shown at their most intense, respective x-y planes ($z = 2.7$ nm for electron, $z = 0.3$ nm for hole). The electron is less confined than our studies with a coarser (0.6 nm) grid and spatial averaging. Calculations of $B$, $C_{eeh}$, $C_{hhe}$ uphold our main results of Auger processes increasing by a larger factor than radiative rates, resulting in alloy fluctuations decreasing internal quantum efficiency as compared to virtual-crystal alloys.

We compare weighted-averaged overlaps for fluctuating alloys in the fine-grid study (Table SI) to the coarse-grid results (Table SII), and find that most overlaps increase by factors of 1.2-2.7 when using finer grids. The exception is the non-polar electron-electron-hole overlap, for which the coarse-grid study predicts a larger overlap by a factor of 1.4. The physical origin of the exception is the reduced localization of electrons in the finer-grid study, which becomes particularly important for eeh Auger recombination (since it involves two less-localized electron states) and for the non-polar direction (for which the carrier localization by fluctuations is not overshadowed by the carrier separation by polarization fields).

TABLE SI. Weighted-averaged overlap results of $In_{0.15}Ga_{0.85}N$ QW study with finer (0.3 nm) grid.

|  | Polar | | | Non-polar | | |
| --- | --- | --- | --- | --- | --- | --- |
|  | $\overline{F^2_{eh}}$ | $\overline{F^2_{eeh}}$ | $\overline{F^2_{hhe}}$ | $\overline{F^2_{eh}}$ | $\overline{F^2_{eeh}}$ | $\overline{F^2_{hhe}}$ |
| Fluctuating alloy | 0.0350 | 0.0084 | 0.0537 | 0.6449 | 0.8632 | 1.1238 |
| Virtual crystal | 0.0079 | 0.0007 | 0.0043 | 0.3533 | 0.3253 | 0.1446 |
| Fluctuating/ virtual crystal ratio | 4.4 | 12.0 | 12.5 | 1.8 | 2.7 | 7.8 |

TABLE SII. Weighted-averaged overlap results of original In$_{0.15}$Ga$_{0.85}$N QW study with 0.6 nm grid and spatially averaged alloy composition.

|  | Polar | | | Non-polar | | |
| --- | --- | --- | --- | --- | --- | --- |
|  | $\overline{F_{eh}^2}$ | $\overline{F_{eeh}^2}$ | $\overline{F_{hhe}^2}$ | $\overline{F_{eh}^2}$ | $\overline{F_{eeh}^2}$ | $\overline{F_{hhe}^2}$ |
| Fluctuating alloy (averaged for 10 distributions) | 0.0175 | 0.0068 | 0.0199 | 0.4812 | 1.2119 | 0.5662 |
| Virtual crystal | 0.0114 | 0.0012 | 0.0068 | 0.3572 | 0.3457 | 0.1552 |
| Fluctuating/ virtual crystal ratio | 1.5 | 5.4 | 2.9 | 1.3 | 3.5 | 3.7 |

We also observe that the non-polar virtual-crystal weighted-averaged overlaps agree very well between the two studies, while the polar virtual-crystal fine-grid overlaps are smaller than those of the original study. This is because the increased resolution in the out-of-plane direction for the polar case allows the spatial separation of electrons and holes from the polarization field to increase slightly (approximately 2.4 nm instead of 2.3 nm), which reduces the net overlap.

By examining the net effect of fluctuations in the fine-grid study compared to the original coarse-grid results, we find that the coarser grids represent the physics and qualitative behavior very well. In all cases examined with the fine-grid calculations, we observe that fluctuations significantly increase overlaps as compared to virtual-crystal alloys, as also observed with coarser grids and spatial averaging. Because the coarser-grid calculations predict smaller overlaps for fluctuating alloys and larger overlaps for virtual-crystal alloys, they slightly underestimate the net effect of alloy fluctuations and carrier localization on the recombination behavior of InGaN quantum wells compared to calculations using finer meshes. Importantly, the fine-grid study predicts an even greater efficiency reduction of the internal quantum efficiency due to alloy fluctuations: 9% (12%) for the polar (non-polar) orientations for the fine-grid study as compared to 4% (9%) for the same carrier density, $10^{19}$ cm$^{-3}$, in the coarse-grid study.

In summary, fine-grid calculations without spatial alloy-composition averaging demonstrate the same physics, qualitative trends, and even greater quantitative efficiency reduction due to alloy fluctuations. Using a 0.6 nm grid and spatially-averaged alloy distribution is a valid method for studying the physics of alloy fluctuations on the InGaN quantum-well IQE.

## IV. DISCUSSION OF EXPERIMENTAL COMPOSITION CHARACTERIZATION

While atom probe tomography (APT) is a very powerful characterization technique, the atomic detection efficiencies are lower than 100%, which may affect the spatial behavior of fluctuations.[12] Hence, results obtained with APT for InGaN[2,4] may depend on experimental and reconstruction parameters.[13,14] Transmission electron microscopy is another useful technique for providing local composition information but may also produce experimental artifacts since radiation damage from overexposure produces localized strain contrast and the artificial appearance of clusters.[2,5,15] Due to the limitations of each technique, it is important to use complementary characterization techniques to verify distribution results. Rigutti *et al.* used effective mass calculations to correlate APT and micro-photoluminescence results for AlGaN, allowing them to statistically correct the composition measured using APT.[12,16] Additionally, for InGaN QWs, Mehrtens *et al.* found scanning transmission electron microscopy (STEM) and APT composition profiles to be in agreement.[17] We note that our calculations did not rely on APT data, but we instead theoretically generated statistically random atomic distributions that agree with experimental distributions. Deviations from a completely random distribution (e.g., In clustering) cause further localization of carriers and exacerbate the droop and green-gap problems. We thus expect that our qualitative results also hold for other disordered atomic distributions that deviate from complete randomness.

## V. DISCUSSION OF EXCITONS

Excitons were not included in this work since they dissociate at the high carrier densities where droop occurs,[18] or at temperatures near room temperature.[19] However, our results at low temperatures and low carrier concentrations, for which excitons have been observed,[19] need to be reexamined with higher-order calculations that account for excitonic effects. Since both radiative and Auger recombination involve one free electron-hole pair recombination (which needs to be updated with the corrected excitonic wave functions), we expect that, to lowest order of approximation, exciton formation increases both the radiative and Auger rates in a similar fashion. Hence, we do not anticipate that our qualitative conclusions regarding the localization effects on the rates and efficiency will be affected by the inclusion of excitonic corrections.

[1] D. Watson-Parris, M.J. Godfrey, P. Dawson, R.A. Oliver, M.J. Galtrey, M.J. Kappers, and C.J. Humphreys, Phys. Rev. B **83**, 115321 (2011).
[2] S.E. Bennett, D.W. Saxey, M.J. Kappers, J.S. Barnard, C.J. Humphreys, G.D. Smith, and R.A. Oliver, Appl. Phys. Lett. **99**, 021906 (2011).
[3] R. Shivaraman, Y. Kawaguchi, S. Tanaka, S.P. DenBaars, S. Nakamura, and J.S. Speck, Appl. Phys. Lett. **102**, 251104 (2013).
[4] J.R. Riley, T. Detchprohm, C. Wetzel, and L.J. Lauhon, Appl. Phys. Lett. **104**, 152102 (2014).
[5] M.J. Galtrey, R.A. Oliver, M.J. Kappers, C.J. Humphreys, D.J. Stokes, P.H. Clifton, and A. Cerezo, Appl. Phys. Lett. **90**, 061903 (2007).
[6] T.-J. Yang, R. Shivaraman, J.S. Speck, and Y.-R. Wu, J. Appl. Phys. **116**, 113104 (2014).
[7] Y. Takagiwa, Y. Pei, G. Pomrehn, and G. Jeffrey Snyder, APL Mater. **1**, 011101 (2013).
[8] B. Delley and E.F. Steigmeier, Appl. Phys. Lett. **67**, 2370 (1995).
[9] K.K. Nanda, F.E. Kruis, H. Fissan, and S.N. Behera, J. Appl. Phys. **95**, 5035 (2004).
[10] B.K. Ridley, *Quantum Processes in Semiconductors* (Oxford : Clarendon Press ; New York : Oxford University Press, 1982).
[11] E. Kioupakis, D. Steiauf, P. Rinke, K.T. Delaney, and C.G. Van de Walle, Phys. Rev. B **92**, 035207 (2015).
[12] L. Rigutti, L. Mancini, D. Hernàndez-Maldonado, W. Lefebvre, E. Giraud, R. Butté, J.-F. Carlin, N. Grandjean, D. Blavette, and F. Vurpillot, J. Appl. Phys. **119**, 105704 (2016).
[13] T.F. Kelly and D.J. Larson, Annu. Rev. Mater. Res. **42**, 1 (2012).
[14] A. Devaraj, D.E. Perea, J. Liu, L.M. Gordon, T.J. Prosa, P. Parikh, D.R. Diercks, S. Meher, R.P. Kolli, Y.S. Meng, and S. Thevuthasan, Int. Mater. Rev. **5**, 1 (2017).
[15] T.M. Smeeton, M.J. Kappers, J.S. Barnard, M.E. Vickers, and C.J. Humphreys, Appl. Phys. Lett. **83**, 5419 (2003).
[16] L. Rigutti, L. Mancini, W. Lefebvre, J. Houard, D. Hernàndez-Maldonado, E. Di Russo, E. Giraud, R. Butté, J.-F. Carlin, N. Grandjean, D. Blavette, and F. Vurpillot, Semicond. Sci. Technol. **31**, 095009 (2016).
[17] T. Mehrtens, M. Schowalter, D. Tytko, P. Choi, D. Raabe, L. Hoffmann, H. Jönen, U. Rossow, A. Hangleiter, and A. Rosenauer, Appl. Phys. Lett. **102**, 132112 (2013).
[18] D. Bayerl, S.M. Islam, C.M. Jones, V. Protasenko, D. Jena, and E. Kioupakis, Appl. Phys. Lett. **109**, 241102 (2016).
[19] M. Shahmohammadi, W. Liu, G. Rossbach, L. Lahourcade, A. Dussaigne, C. Bougerol, R. Butté, N. Grandjean, B. Deveaud, and G. Jacopin, Phys. Rev. B **95**, 125314 (2017).